# A Pixel-based Encryption Method for Privacy-Preserving Deep Learning Models

Ijaz Ahmad[1] and Seokjoo Shin[*]
[1,*]Dept. of Computer Engineering
Chosun University
Gwangju, 61452 South Korea
[1]ahmadijaz@chosun.kr, [*]sjshin@chosun.ac.kr (Corresponding author)

*Abstract*—In the recent years, pixel-based perceptual algorithms have been successfully applied for privacy-preserving deep learning (DL) based applications. However, their security has been broken in subsequent works by demonstrating a chosen-plaintext attack. In this paper, we propose an efficient pixel-based perceptual encryption method. The method provides a necessary level of security while preserving the intrinsic properties of the original image. Thereby, can enable deep learning (DL) applications in the encryption domain. The method is substitution based where pixel values are XORed with a sequence (as opposed to a single value used in the existing methods) generated by a chaotic map. We have used logistic maps for their low computational requirements. In addition, to compensate for any inefficiency because of the logistic maps, we use a second key to shuffle the sequence. We have compared the proposed method in terms of encryption efficiency and classification accuracy of the DL models on them. We have validated the proposed method with CIFAR datasets. The analysis shows that when classification is performed on the cipher images, the model preserves accuracy of the existing methods while provides better security.

## I. INTRODUCTION

Training deep learning (DL) models requires cutting-edge technology and high computational resources, which is sometimes unaffordable by small organizations and independent researchers in underprivileged regions. Cloud-based computing is one of the cost-effective solutions to the aforementioned problem. It is becoming popular these days for training DL models instead of using local servers, thanks to the cloud computing platforms like Google cloud, Amazon Web Services, Microsoft Azure etc. However, the data transfer over public networks is always at a risk of being leaked. A simple solution is to encrypt the data before transmission. The traditional full encryption methods like RSA, AES have proven to be the most secure options for image data; however, the encrypted images need to be decrypted in order to perform image processing and computer vision tasks on them. The data reveal to the third-parties is acceptable in certain situations; however, when the images are privacy-sensitive like medical image data, financial data, surveillance data etc., then the traditional encryption algorithms are inefficient. Alternative solutions have been proposed like Homomorphic encryption and multi-party computations, which are computationally expensive and the DL models accuracy suffers [1].

In the recent years, perceptual encryption (PE) algorithms have been proposed to enable privacy-preserving machine learning. The algorithms perform encryption in such a way that the images are visually unrecognizable but can still be interpreted by machines. The block-based PE algorithms proposed in [2] have been applied to traditional ML methods like support vector machine (SVM) [1] but have not been considered for DL models. On the other hand, pixel-based encryption algorithms have been successfully applied to DL models in [3]–[5]. A pixel-based PE method called Learnable Encryption (LE) has been proposed by [3], which is successful in preserving the accuracy of the DL models. However, it is vulnerable to chosen-plaintext attack demonstrated in [6]. The security of LE has been improved by extended learnable encryption (ELE) method proposed in [5]; however, the accuracy of DL is reduced significantly. Alternatively, a pixel-based encryption method proposed in [4], we refer to it as PBE, has better security (i.e., keyspace) than LE and preserves the DL accuracy. However, it is also vulnerable to chosen-plaintext attack shown in [6]. It can be seen that there is a necessity for an encryption algorithm that is secure, computationally inexpensive and does not degrade the accuracy of DL models.

In this paper, we propose an efficient pixel-based perceptual encryption algorithm. The method provides the necessary level of security while preserving the intrinsic properties of the original image. Thereby, images encrypted with the proposed method can enable privacy-preserving DL models. The encryption is substitution based where pixel values are XORed with a key generated by a chaotic map. We have used logistic map for the substitution for its low complexity [7]. In addition, we have used a second key to shuffle the chaotic map in order to compensate for any inefficiency caused by the logistic maps. Both keys together are used as encryption and decryption keys. The proposed algorithm is independent of the DL models; therefore, it can be used with any DL algorithm.

The rest of the paper is organized as: in the next section we present the conventional and proposed perceptual encryption methods to generate cipher images. Then, we present our experimental setup and results in Section III. Finally, Section IV concludes the paper and gives the future research direction.

## II. PROPOSED METHOD

In this section, we first discuss the conventional pixel-based PE methods and highlight their limitations. Then, we present the proposed PE method. In Section III, we carry out comparisons with all the LE [3], ELE [5] and PBE [4] but here we consider the last one in more detail.

### A. Pixel-based Encryption Method [4]

To encrypt a color image $I_{(x,y,c)}$, where the pair $(x, y)$ is the position of a pixel in the color channel $c$, the PBE method consists of the following two steps:

1) Apply negative and positive transformation to each pixel $p_{(x,y,c)}$ of the image by using a random binary key $K_c$ as:

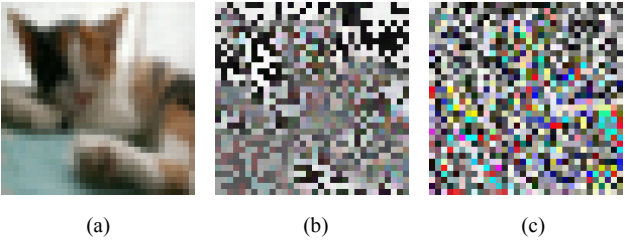

Fig. 1. Examples of inputs to deep learning model. (a) Original image. (b) and (c) are cipher images obtained from pixel-based encryption method proposed in [4] and the proposed method, respectively.

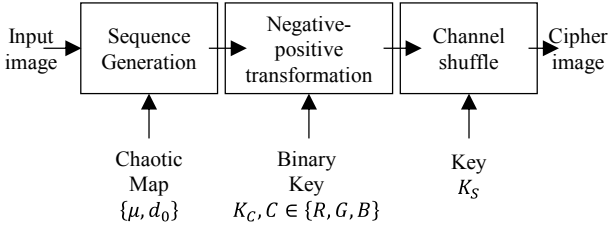

Fig. 2. Proposed pixel-based perceptual encryption method.

$$p'_{(x,y,c)} = \begin{cases} p_{(x,y,c)}, & k=0 \\ p_{(x,y,c)} \oplus (2^L - 1), & k=1 \end{cases} \quad (1)$$

where $L$ is the levels of intensity and $k$ is an entry in the key $K_c$. The key $K_c$ is uniformly distributed.

2) An optional step of shuffling the color channels of each pixel by using key $K_s$ given in Table I.

For an image with $n$ pixels, the keyspace of PBE is a product of $K_C$ and $K_S$ and is given as,

$$K_{PBE} = 2^{3n} \cdot 6^n \quad (2)$$

The keyspace is large enough to resist brute-force attacks. Fig. 1. b. shows an example cipher image generated by the PBE method. It can be seen that no visual information is visible. However, [6] has demonstrated a successful chosen-plaintext attack against the PBE. As oppose to the attack in [6] where an image is recovered, we have implemented a variation of the attack to recover the key as shown in Algorithm 1. We assumed that no color shuffle has been applied but it can be modified easily as in [6]. The algorithm returns key $K_C$ for each color channel and $C \in \{R, G, B\}$.

### B. Proposed Perceptual Encryption Method

The vulnerability of [3], [4] stems from the choice of binary key used in the negative and positive transformation step i.e., the xor operation of the pixel value with a single $(2^L - 1)$ value. Therefore, in the proposed method, we use a sequence of values ranging from 0 to $(2^L - 1)$. For this purpose, we utilize logistic chaotic map [8] given by

$$d_{i+1} = \mu d_i (1 - d_i) \quad (3)$$

where, $\mu$ is the control parameter and the system is chaotic for $\mu \in [3.57, 4]$. Both parameter $\mu$ and the initial value $d_0$ serve as the key to generate the map. The logistic map has Lyapunov exponent 0.0012 and 0.00693 for $\mu = 3.57$ and $\mu = 4$, respectively [7]. The proposed algorithm is illustrated in Fig. 2., and consists of the following steps:

**Algorithm 1:** Chosen-Plaintext attack
**Input:** Input image size $X \times Y$; level of intensities $L$
**Output:** Key $K$ of size $X \times Y \times 3$.

1. Initialize key of size $X \times Y \times 3$ as $K \leftarrow 0$
2. Choose a positive integer $a$ such that $a < L$ and $a \neq (a \oplus L)$;
3. Initialize image of size $X \times Y \times 3$ as $E \leftarrow a$
4. Encrypt $E \leftarrow \text{Enc}(E)$
5. **foreach** $C \in \{R, G, B\}$
6.    **foreach** $p_C \in \text{Enc}(E_c)$ **do**
7.      **if** $p_{(x,y)} = a \oplus (2^L - 1)$ **then**
8.        $K_{(x,y,C)} \leftarrow 1$;
9. **return** $K$

**Algorithm 2:** Generate sequence
**Input:** Control parameters $miu, d$; intensity levels $L$; image size $X \times Y$
**Output:** Sequence $S$

1. **for** $i \leftarrow 1:R$
2.    $d \leftarrow miu * d * (1-d)$
3. **for** $i \leftarrow 1: X \times Y$
4.    $d \leftarrow miu * d * (1-d)$
5.    $S[i] \leftarrow (\lfloor (d/2) * 10^{14} \rfloor) \% L$
6. **return** $S$

TABLE I. COLOR CHANNEL SHUFFLE KEY

| Key $K_S$ | Red (R) | Green (G) | Blue (B) |
|---|---|---|---|
| 0 | R | G | B |
| 1 | R | B | G |
| 2 | G | R | B |
| 3 | G | B | R |
| 4 | B | R | G |
| 5 | B | G | R |

1) Generate a sequence $S$ for negative and positive transformation step as,

$$S_i = \left( \left\lfloor \left( \frac{d}{2} \right) \times 10^{14} \right\rfloor \right) \% L \quad (4)$$

where, $L = 256$ is the level of intesities, and $d$ is the state of the map. The sequence generation is demonstrated in Algorithm 2. In Line 1, we discard $R = 3000$ initital values of the sequence in order to avoid harmful effect of the map.

2) Perform negative and positive transformation of each pixel $p_{(x,y,c)}$ of the image with sequence $S$ by using a random binary key $K_c$ as:

$$p'_{(x,y,c)} = \begin{cases} p_{(x,y,c)}, & k=0 \\ p_{(x,y,c)} \oplus S_{(i,c)}, & k=1 \end{cases} \quad (5)$$

where $k$ is an entry in the key $K_c$. The key $K_c$ is uniformly distributed and choose randomly the pixels to be modified. Therefore, not all the values of the sequence are used. The election add randomness to the logistic map in order to compensate for any inefficiency caused by the map.

TABLE II. CLASSIFICATION ACCURACY (%) OF PRIVACY-PRESERVING DEEP LEARNING MODEL.

| Dataset | Plain Image | Learnable Encryption (LE) | Pixel-based Encryption (PBE) | Extended Learnable Encryption (ELE) | Proposed |
|---|---|---|---|---|---|
| CIFAR10 | 96.70 | 94.94 | 94.27 | 67.10 | 94.58 |
| CIFAR100 | 83.59 | 78.52 | 78.96 | 43.05 | 78.39 |

3) Apply color channel shuffling for each pixel using a randomly generated key $K_S$ as given in Table I.

For an image with $n$ pixels, the keyspace of the proposed method is a product of $S, K_C$ and $K_S$ and is given as

$$K_{PBE} = 256^{3n} \cdot 2^{3n} \cdot 6^n \quad (6)$$

The keyspace is large enough to resist brute-force attack. It can be seen that the use of a sequence instead of single value makes the propose algorithm resistance to the chosen-plaintext attack demonstrated in Algorithm 1. For visual analysis, Fig. 1. (c) shows a cipher image generated by the proposed pixel-based PE algorithm.

## III. RESULTS AND DISCUSSION

In this section, we first present our experimental setup. Then, we show the performance of a DL based classifier on the plain images as well as encrypted images obtained from Learnable Encryption (LE), Extended Learnable Encryption (ELE), Pixel-based Encryption (PBE) and the proposed encryption methods.

We have used PyramidNet [9] with ShakeDrop regularization [10] as the classifier. The same training parameters were used as of [5]. The experiments have been performed on CIFAR10 and CIFAR100 [11]. Both datasets consists of 60k tiny color images with resolution of 32x32. The datasets have 50k training images and 10k test images. CIFAR10 has 10 classes and 6k images per class, while CIFAR100 has 100 classes and 600 images per class. In our experiments, we have used 10% of training images as our validation images. In our experiments, we first encrypt training and testing datasets before input to the model. We have used the same key for both datasets. Table II summarizes the classification accuracy of the classifier on the test dataset. The plain images, and the cipher images based on LE [3], ELE [5], PBE [4], and the proposed methods are compared. The accuracy reported is in percentage, higher accuracy means better performance. The ELE accuracy is the worst and the rest of three methods have comparable accuracy. For the proposed method, a degradation of 2% and 5% in accuracy occurs on CIFAR10 and CIFAR100 datasets compared to the plain images, respectively.

## IV. CONCLUSION AND FUTURE WORK

In this paper, we proposed a pixel-based perceptual encryption algorithm to enable privacy-preserving deep learning (DL) applications. We explored the idea of modifying the pixels values using a sequence of values instead of a single value as in the existing methods. By doing so, the proposed method is robust against the chosen-plaintext attack. The larger keyspace, i.e., the sequence length, key for substitution stage and key for color channel shuffling, makes the algorithm secure against brute-force attacks. In addition, when a DL based classification is performed on the cipher images, the model preserves the accuracy as that of the existing methods; however, there is a slight degradation as compared to the plain images.

The encryption method is based on logistic chaotic map, which have some inefficiency. We have taken a measure to compensate for the inefficiency; however, in future we are interested to secure the algorithm by adapting more secure chaotic maps.

ACKNOWLEDGMENT

This research is supported by Basic Science Research Program through the National Research Foundation of Korea (NRF) funded by the Ministry of Education (NRF-2018R1D1A1B07048338).